08/01/2016

# High current, high efficiency graded band gap perovskite solar cells

Onur Ergen, [1,3,4] S. Matt Gilbert[1, 3,4], Thang Pham[1, 3,4] , Sally J. Turner, [1,2,4], Mark Tian Zhi Tan [1], Marcus A. Worsley[5], and Alex Zettl[1, 3,4,*]

[1]Department of Physics, University of California at Berkeley, Berkeley, California 94720, USA
[2]Department of Chemistry, University of California at Berkeley, Berkeley, California 94720, USA
[3]Materials Sciences Division, Lawrence Berkeley National Laboratory, Berkeley, California 94720, USA
[4]Kavli Energy Nanosciences Institute at the University of California, Berkeley, and the Lawrence Berkeley National Laboratory, Berkeley, California 94720, USA
[5] Physical and Life Sciences Directorate, Lawrence Livermore National Laboratory, Livermore, CA 94550, USA

*Author to whom correspondence should be addressed: azettl@berkeley.edu

**Organic-inorganic halide perovskite materials have emerged as attractive alternatives to conventional solar cell building blocks. Their high light absorption coefficients and long diffusion lengths suggest high power conversion efficiencies (PCE),[1-5] and indeed perovskite-based single band gap and tandem solar cell designs have yielded impressive performances.[1-16] One approach to further enhance solar spectrum utilization is the graded band gap, but this has not been previously achieved for perovskites. In this study, we demonstrate graded band gap perovskite solar cells with steady-state conversion efficiencies averaging 18.4%, with a best of 21.7%, all without reflective coatings. An analysis of the experimental data yields high fill factors of ~75% and high short circuit current densities up to 42.1 mA/cm$^2$. These cells, which are based on a novel architecture of two perovskite layers ($MASnI_3$ and $MAPbI_{3-x}Br_x$), incorporating GaN, monolayer hexagonal boron nitride, and graphene aerogel, display the highest efficiency ever reported for perovskite solar cells.**

Organic-inorganic perovskite solar cells are typically prepared in a single band gap configuration, where an absorber layer ($ABX_3$, A=$CH_3NH_3$(MA);B=Pb ,Sn; and X=Cl, Br, I) is sandwiched between an electron injection layer (ETL) and a hole transport layer (HTL).[1-5] Following significant effort in optimizing interface layers to control the carrier dynamics, power conversion efficiencies (PCE) for this design, for a single cell, have topped 19% (one study reported 19.3%[2] and another 20.1%[3]) for lead based perovskite solar cells. Recently, cesium containing triple cation perovskite solar cells, also lead based, have reached 21.1% PCE.[4] However, average device efficiencies are reported as no more than ~17%.[2-4] In addition, due to the toxicity of lead in the absorber layer, lead free tin halide perovskite solar cells have gained tremendous importance. However, lead free cells do not display such high photovoltaic performances (less than 7%) due to chemical instability.[5] The tunable band gap of methylammonium-lead-halide has also led researchers to construct multijunction tandem cells which aim to maximize the solar irradiative spectrum.[6-12] In these tandem cells, the perovskite layer can be integrated with crystalline silicon (c-Si) and copper indium gallium selenide (CIGS). However, the tandem cell requires complex electrical coupling and interconnection between the perovskite sub-cells, which generates electron-hole recombination centers. In spite of numerous proposals for band gap engineering of perovskite layers by replacing the metal cations, varying the composition of halide ions, or altering the moisture content, only one report has emerged of a successful perovskite/perovskite two terminal tandem cell[12], with a PCE of 7%. An appealing alternative is the perovskite-based graded band gap solar cell, for which, in principle, the electron-hole collection efficiency can be enhanced considerably, resulting in acceptable open circuit output voltage and very large output current. In contrast to tandem cells, complex interconnections and current coupling are not needed in this architecture. Despite these advantages, a functioning perovskite-based graded band gap solar cell has proved elusive, likely due to excessive cation mixing.

Here we report high efficiency graded band gap perovskite solar cells with very large current outputs. We fabricate mixed halide double-layer perovskite devices (layer 1:$CH_3NH_3SnI_3$ and layer 2: $CH_3NH_3PbI_{3-x}Br_x$) in order to create a graded band gap. Perovskite layers are deposited on a heavily doped gallium nitride (GaN) substrate, which in turn serves as an electron injection layer. A monolayer of hexagonal boron nitride (h-BN) is used as a cationic diffusion barrier and adhesion promoter between these two layers, in addition to its excellent electron tunneling properties. Moreover, we manipulate the carrier transport properties in the spiro-OMeTAD based HTL by incorporating a graphene aerogel (GA). The architecture is robust and the cells reliably produce very large current densities up to 45mA/cm$^2$ with average PCEs of 18.41%, with the highest steady-state PCE topping 21.7% (freshly illuminated cells display PCEs of nearly 26%).

Fig.1a shows schematically the stacked architecture of the graded band gap perovskite solar cell. The positions of the key h-BN and GA components are clearly indicated. Briefly, the cells are fabricated as follows (full details in the methods section): solution processing is employed using mixed halide perovskite solutions ($CH_3NH_3SnI_3$ and $CH_3NH_3PbI_{3-x}Br_x$) on plasma etched GaN at room temperature. A monolayer of h-BN is sandwiched between these two absorbers to prevent possible cation mixing. A hole-transporting layer (spiro-OMeTAD based incorporated into GA) and a final current collecting layer (Au) are then deposited. Contact to the GaN layer is via a stacked Ti/Al/Ni/Au finger electrode. A cross sectional scanning electron microscope (SEM) image of the device is shown in Fig. 1b.

Fig. 2 shows photoluminescence (PL) spectra of typical devices and their performances under constant illumination. Photocurrent generation for a complete device (with h-BN and GA modifications) begins in the range of ~600nm to ~1300nm. The presence of broad and multiple peaks implies the formation of a graded band gap. Using Sn as an active metal cation in the first perovskite layer provides a narrow band gap in the 1.2eV to 1.5eV range. [13-18] However, Sn containing semiconductors can be strain sensitive, which can cause further narrowing of the bandgap.[19] The strain at the GaN/Perovskite interface can result in significant energy band gap shifts due to splitting of valance band degeneracy which leads to the lowest bandgap of our device, around 1eV, and results in an enhanced photocurrent generation up to 1250nm. Replacing Sn with Pb in the second layer facilitates a larger band gap between 1.5 eV to 2.2 eV by varying halide composition. [20-23] The experimental evidence of this effect is also shown in the absorption and steady-state photoluminescence spectra (Fig. S1). The red shifts of the luminescent peaks (Fig. S1) are due to this gradual increase in iodide fraction. An energy dispersive x-ray spectroscopy (EDAX) line scan analysis also shows characteristic features of cationic diffusion and confirms the variation in iodide concentration (Fig. S5b).

Fig. 3a shows external quantum efficiency (EQE) measurements for devices with and without GA and h-BN modifications. The data show significant differences in spectral response. For samples with h-BN and GA, the response extends up to ~1400nm (increasing the theoretical short circuit limit to 50mA/cm$^2$). The long wavelength absorption, higher than 1250nm, arises due to an extra PL peak in the near infrared region (NIR) which appears only under constant illumination. This light induced peak forms at ~1300nm and broadens with increased light intensity (Fig. S11). This peak can be attributed to two main effects: a possible defect induced lattice absorption or a free carrier accumulation which results in charge screening at the band edges; thus, the bandgap is further reduced. This band gap narrowing is independent of strain induced bandgap lowering and arises only under illumination.[24,25] The EQE data also clearly indicate excellent light trapping properties due to the textured surface of GaN caused by residual etching (Fig. S14). The cells without h-BN and GA modifications exhibit poor spectral response at long wavelengths which progressively decreases over time. This confirms the critical importance of h-BN and GA modifications. In the EQE data, the cells without GA also display compositional fluctuations due to ionic motion with more incomplete collection of photo-generated charge. However, the high surface area of the GA helps to reduce fluctuations and improve collection efficiency. The current density-voltage (J-V) characteristics of these devices are shown in Fig. 3b. J-V parameters are measured (2400 Series Source Meter Keithley Instruments) under AM 1.5 illumination at an intensity of $1000 Wm^{-2}$. The measured short circuit current density ($J_{sc}$) ranges from ~25 $mA/cm^2$ to ~45 $mA/cm^2$. These large $J_{sc}$s are record-setting for perovskite solar cells. We suspect carrier multiplications, such as impact ionization and multi-exciton formation, might also play an important role in Jsc improvement due to a strong built-in electric field in the device.[26,27] The cells with GA and h-BN modifications show the highest current output and efficiency. Other current-voltage trends can also be seen in Fig. 3b. The graded band gap formation (cells with GA and h-BN modifications) provides an effective built-in electric field, which also enhances the electron hole collection efficiency but necessarily lowers the open circuit voltage ($V_{oc}$). $V_{oc}$ ranges from ~0.64V to ~0.9V for these cells and is limited by the lowest bandgap of the device.

We have found that perovskite-based solar cells have time-dependent performance characteristics. Freshly illuminated cells tend to have higher PCE, for example, than cells that have been illuminated for more than a few minutes. Fig. 4a shows this trend for a given graded band gap perovskite cell. Within the first two minutes of illumination and characterization, the PCE is between 25% and 26%. After approximately 5 minutes, the cell reaches a "steady state", with stable performance (in this case a PCE of 20.8%). In this report, the performance characteristics we quote are for the steady state (we note that, regrettably, the perovskite solar cell literature is not standardized in this practice, with many quoted performance values likely valid only for the early "transient" state). Fig. 4b shows a histogram for all 40 graded band gap perovskite cells measured, having the configuration shown in Fig. 1. The average steady state PCE over all devices is 18.4%. The average fill factor (FF) for the same set of devices is 72% (not shown in Fig. 4b), and the cells consistently exhibit similar characteristics between reverse and forward sweep directions (Fig. S9). The measured solar cell parameters of our best graded band gap cell in the steady state are $J_{sc}$=42.1 mA/cm$^2$ , $V_{oc}$= 0.688 V, FF=0.75, and PCE = 21.7% (mean value of PCE=21.66%, surface area of 0.07cm$^2$). The current density-voltage (J-V) characteristics of this cell are shown in Fig. 4c and integrated spectral response can be seen in Fig. S13. These, the highest efficiency cells, have Jsc=42.1 mA/cm$^2$ out of a possible 49.4 mA/cm$^2$ available for a bandgap ~1eV under AM1.5 global illumination.

Table 1 summarizes best performance characteristics of perovskite solar cells, including single band gap devices (refs. 2 and 3), cesium containing triple cation cells (ref. 4), $MASnI_3$ cells (ref. 5), perovskite-perovskite tandem cell (ref. 14), and graded band gap cells (this work). As expected, the tandem cells have the highest $V_{oc}$, but at the expense of a disappointingly low PCE. The graded band gap cells display the highest $J_{sc}$ and highest PCE.

We now discuss further the architecture and internal functioning of the graded band gap perovskite solar cells. Devices without a monolayer h-BN layer between the mixed halide double layer perovskites exhibit almost no graded bandgap characteristics at any time. Moreover, such devices consistently exhibit low performance and a rapid photocurrent decrease (Figs. 2 and Fig. 3). This demonstrates that h-BN plays a key role in facilitating the graded band gap function. EDAX line mapping of cells with and without h-BN incorporation can be seen in Figs. S5b and S5c. Sn and Pb concentrations drastically diminish from one layer to another at the h-BN interface, which demonstrates that h-BN acts as a diffusion barrier to prevent cation mixing. Devices with h-BN and GA modifications also exhibit stable electrical characteristics even under constant illumination (Fig. S6). These characteristics are likely due to increased oxidation or segregation of iodide at the interfaces of h-BN and HTM/GA. Furthermore, tin may tend to have stronger bonds with bromine at the h-BN interface, forming an intermediate medium of $CH_3NH_3SnBr_3$.[15,16]

The graphene aerogel (GA) acts as a barrier to moisture ingress (Fig. S2d). The barrier may alleviate moisture penetrating into deeper depths of the absorber layer and help maintain the interface stability. It is well known that humidity exposure, or a decreasing iodide fraction, leads to a wider band gap perovskite up to ~2.4eV (Figs. S2 and Fig. S5). [20-25] EDAX line mapping of the oxygen signature of a perovskite with and without GA shows dramatic differences (Fig. S2d). GA also plays an important role in shaping the crystallinity and morphology of the perovskite film due to its high surface area. The GA modification is critical to obtain highly crystalline and homogeneous perovskite films, and we find that without GA modifications the perovskite films have significantly smaller grain sizes and form isolated perovskite islands rather than continuous films (Figs. S4).[28] Limiting the nucleation of small islands with the GA modification has the benefit of allowing for quick growth and aggregation, promoting large grain sizes (Figs. S3 and S4). Furthermore, the mobility also exhibits a clear dependence on the presence of a GA layer; all of the films with a GA layer exhibit better performance than without it (Fig. S7). GA is a key component in ultra high performance devices.

In summary, graded band gap perovskite photovoltaic cells are prepared successfully with record output current and power efficiency by implementing a new cell architecture. GaN is chosen to replace typical $TiO_2$ ETL to provide a better surface morphology and enhanced electron injection due to its ability to dope heavily. A graphene aerogel makes an excellent barrier layer to moisture ingress and improves hole collection efficiency in the HTL. The aerogel also promotes a more crystalline perovskite structure. Choosing the right metal cation and varying halide anion concentration also successfully establishes band gap tuning of the perovskite absorber layer. The combination of GA and h-BN enables this band gap formation, and this configuration produces cells that are remarkably reproducible and stable.

**Methods:**

**Fabrication:** Commercial GaN on a silicon (Si) wafer was annealed at 650 $^{o}$C for 2h in argon (Ar) environment. The backside of the wafer (silicon surface) was mechanically polished by diamond paste, until the silicon layer was thin. This thin layer was photolithographically masked by a silicon nitride ($Si_3N_4$) film and etched entirely by 45wt % potassium hydroxide (KOH, Sigma-Aldrich) at 110 $^{o}$C for 14h. Then a Ti/Al/Ni/Au (30/100/20/150 nm) stack layer was deposited by e-beam lithography and e-beam evaporation followed by rapid thermal annealing at $850^0C$. The GaN surface was briefly plasma etched to help evenly disperse the perovskite solution. Next, $CH_3NH_3SnI_3$ was spin coated at 4000 r.p.m for 45 s and crystallized at 80 $^oC$. Afterwards a monolayer h-BN was transferred directly onto the prepared substrate[29]. The HTM layer was deposited on a graphene aerogel (GA) by spin coating 2000 r.p.m for 30 s and then left at room temperature for 5 min. Subsequently, $CH_3NH_3PbI_{3-x}Br_x$ was spin coated on this GA/HTM layer at 3000 r.p.m for 30 s and the film crystallized at 60°C. This second layer was gently placed onto the first layer, in the glove box, and annealed at 60°C. Finally, a 75nm thick gold (Au) electrode was evaporated on top of the HTM/GA layer. $CH_3NH_3SnI_3$ and $CH_3NH_3PbI_{3-x}Br_x$ were synthesized according to procedures published in reference *16* and *20*. The HTM layer was prepared according to refs. *9* and *13*. h-BN was prepared as in ref. *29* and the GA sheets were prepared by the gelation of a GO suspension. The aqueous GO suspension (2 wt%) was prepared by ultrasonication. In a glass vial, 3 ml of the GO suspension was mixed with 500 uL (microliter) of concentrated NH4OH (28-30%). The vial was sealed and placed in an oven at 80°C overnight. The resulting wet gel was washed in deionized water to purge $NH_4OH$ followed by an exchange of water with acetone inside the pores. The washed gel then underwent supercritical drying by using $CO_2$ and was converted to the final graphene aerogels by pyrolysis at 1050°C under nitrogen flow.[30]

**Characterization:** The XRD spectra were measured using a Siemens D500 X-ray diffractometer, and EQE measurements were performed using a QEPVSI measurement system (Newport 300 W xenon lamp, 66920, Newport Cornerstone 260 monochromator, and Lock-in amplifier SRS810). UV/VIS absorption spectra were recorded on a PG T80 spectrophotometer in the 1190-1100nm wavelength range at room temperature. Photoluminescence spectra were measured with a modified Renishaw inVia Raman microscope (Ar ion laser 514nm and 488nm and HeNe 633nm). J-V curves were measured using a solar simulator (Newport, 91195A) with a source meter (Keithley 2420) at $1000W/m^2$ AM 1.5 illuminations, and a calibrated Si-reference cell certified by NREL. The voltage sweep was maintained at very slow rate (<$10mV/s^{-1}$), to insure that the device reached its equilibrium and yielded a hysteresis-free response. J-V data were also corrected by the spectral mismatch factor, calculated to be less than 1%. All devices were measured while masking the active area with a metal mask of size 0.07 $cm^2$. All measurements were conducted at room temperature in air.

**Acknowledgements**

The authors thank B. Lechene (Prof. A. Arias group) and D. Hellebusch, S. Hawks, N. Bronstein ( Prof. P. Alivisatos group) for use of the solar simulator, J. Kim and C. Jin (Prof. F. Wang group) for PL measurements and discussions, E. Cardona (Prof. O. Dubon group) for XRD measurements, L. Leppert (Prof. J. Neaton group) for valuable discussions on investigation of band gap alignment, and T. Moiai and K. Emery (National Renewable Energy Laboratory) for valuable technical discussions on calibration, J-V measurements, and EQE measurements. This research was supported in part by the Director, Office of Science, Office of Basic Energy Sciences, Materials Sciences and Engineering Division of the U.S. Department of Energy under Contract No. DE-AC02-05CH11231, which provided for PL measurements under an LDRD award, and, within the $sp^2$-bonded materials program (KC2207), for the design of the experiment and material characterization; the National Science Foundation under Grant 1542741, which provided for photovoltaic response characterization; and by the Office of Naval Research (MURI) under Grant N00014-12-1-1008, which provided for h-BN growth. This work was additionally supported by Lawrence Livermore National Laboratory under the auspices of the U.S. Department of Energy under Contract DE-AC52-07NA27344 through LDRD 13-LW-099 which provided for graphene aerogel synthesis. SMG acknowledges support from the NSF Graduate Fellowship Program.

**Author contributions**

O.E., S.M.G., T.P., S.J.T., and A.Z designed the experiments. O.E., S.M.G., T.P., S.J.T., and M.T.Z.T. carried out experiments. O.E., S.M.G, T.P, S.J.T., M.T.Z.T., and A.Z contributed to analyzing the data. O.E. and A.Z. wrote the paper and all authors provided valuable feedback.

**Competing financial interests**

The authors declare that they have no competing financial interests.

**Table 1** The best device performance values of perovskite solar cells, measured under 1000Wm$^{-2}$ AM 1.5G illuminations.

| Sample | Voc (V) | Jsc (mA/cm$^2$) | FF (%) | PCE (%) | Area (cm$^2$) |
|---|---|---|---|---|---|
| Ref. [2] (Single band gap) | 1.13 | 22.75 | 75.01 | 19.3 | 0.10 |
| Ref. [3] (Single band gap) | 1.059 | 24.65 | 77 | 20.1 | 0.09 |
| Ref. [4] (Cesium containing triple cation perovskite) | 1.147 | 23.5 | 78 | 21.1 | 0.16 |
| Ref. [12] (Perovskite/Perovskite tandem) | 1.89 | 6.61 | 56 | 7 | 0.04 |
| Ref. [5] (Lead free tin halide perovskite solar cell)($MASnI_3$) | 0.88 | 16.8 | 0.42 | 6.4 | 0.06 |
| This Work (Graded band gap) | 0.688 | 42.1 | 75 | 21.7 | 0.07 |

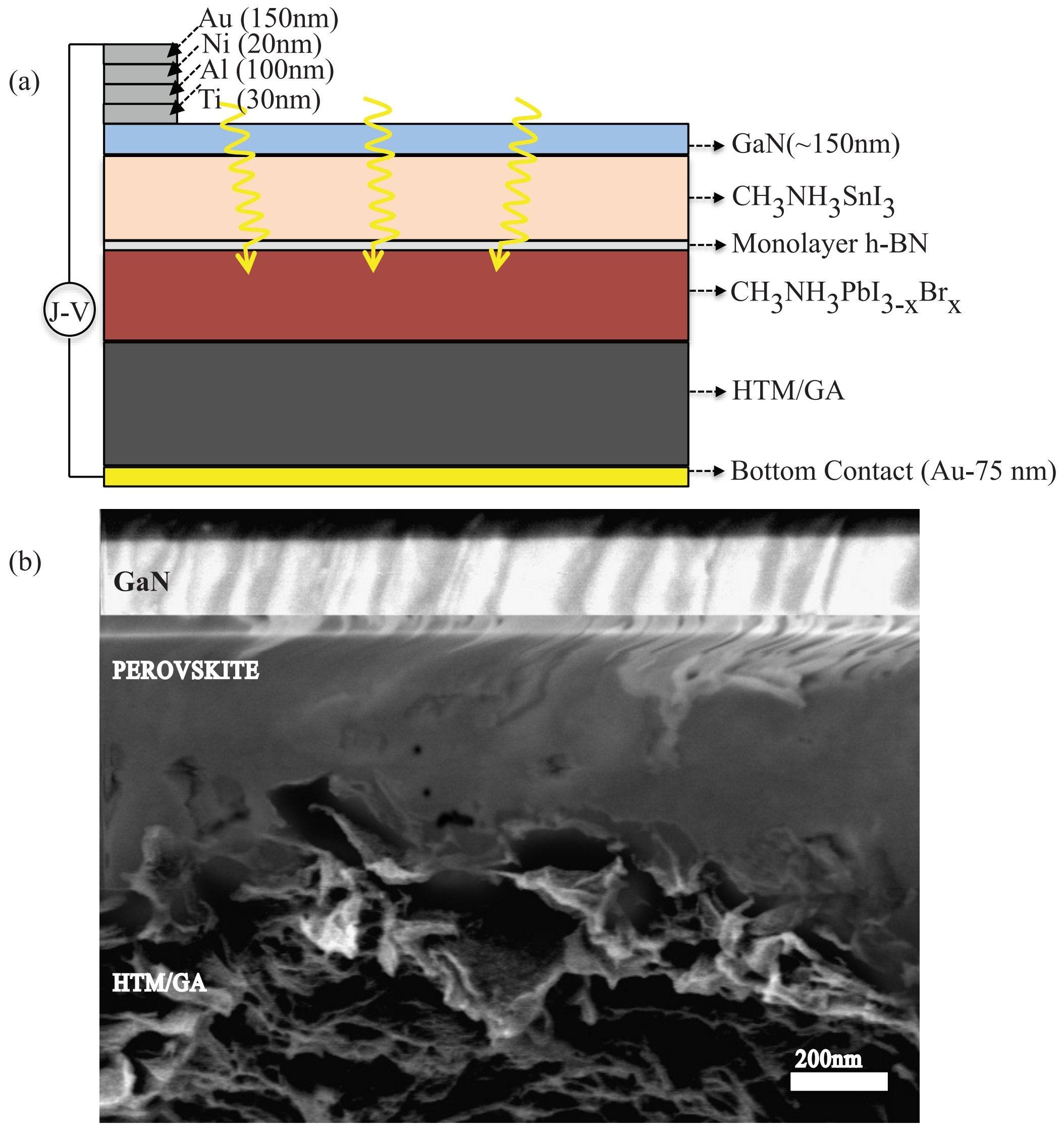

**Figure 1**(a) Schematic of a graded band gap perovskite solar cell. Gallium nitride (GaN), monolayer hexagonal boron nitride (hBN), and graphene aerogel (GA) are key components of the high efficiency cell architecture. (b) Cross-sectional scanning electron microscopy (SEM) image of a representative perovskite device. Monolayer h-BN between perovskite layers is not visible in this SEM image. (Thickness of the $CH_3NH_3SnI_3$ layer is ~150nm and the $CH_3NH_3PbI_{3-x}Br_x$ is ~300nm)

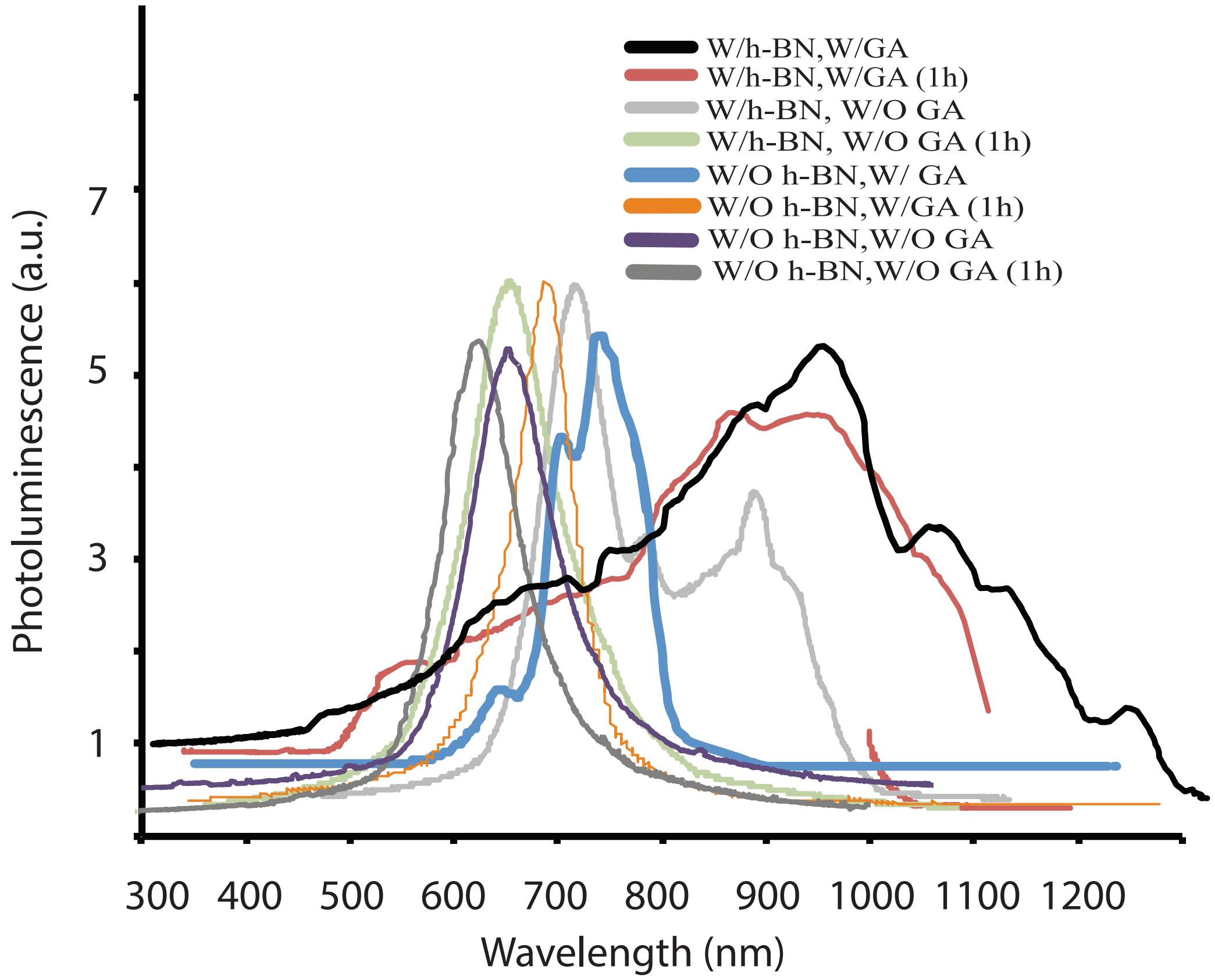

**Figure 2** Photoluminescence (PL) spectra of perovskite cells with (W) and without (W/O) monolayer hBN or graphene aerogel (GA) components. The data are recorded in the steady state regime after a few minutes of constant illumination, or after one hour (1h) of constant illumination. Cells with hBN and GA show significant stability over time. Cells without GA but with h-BN layer exhibit brief graded band gap formation and moderate degradation afterwards. Cells without h-BN exhibit no graded band gap formation.

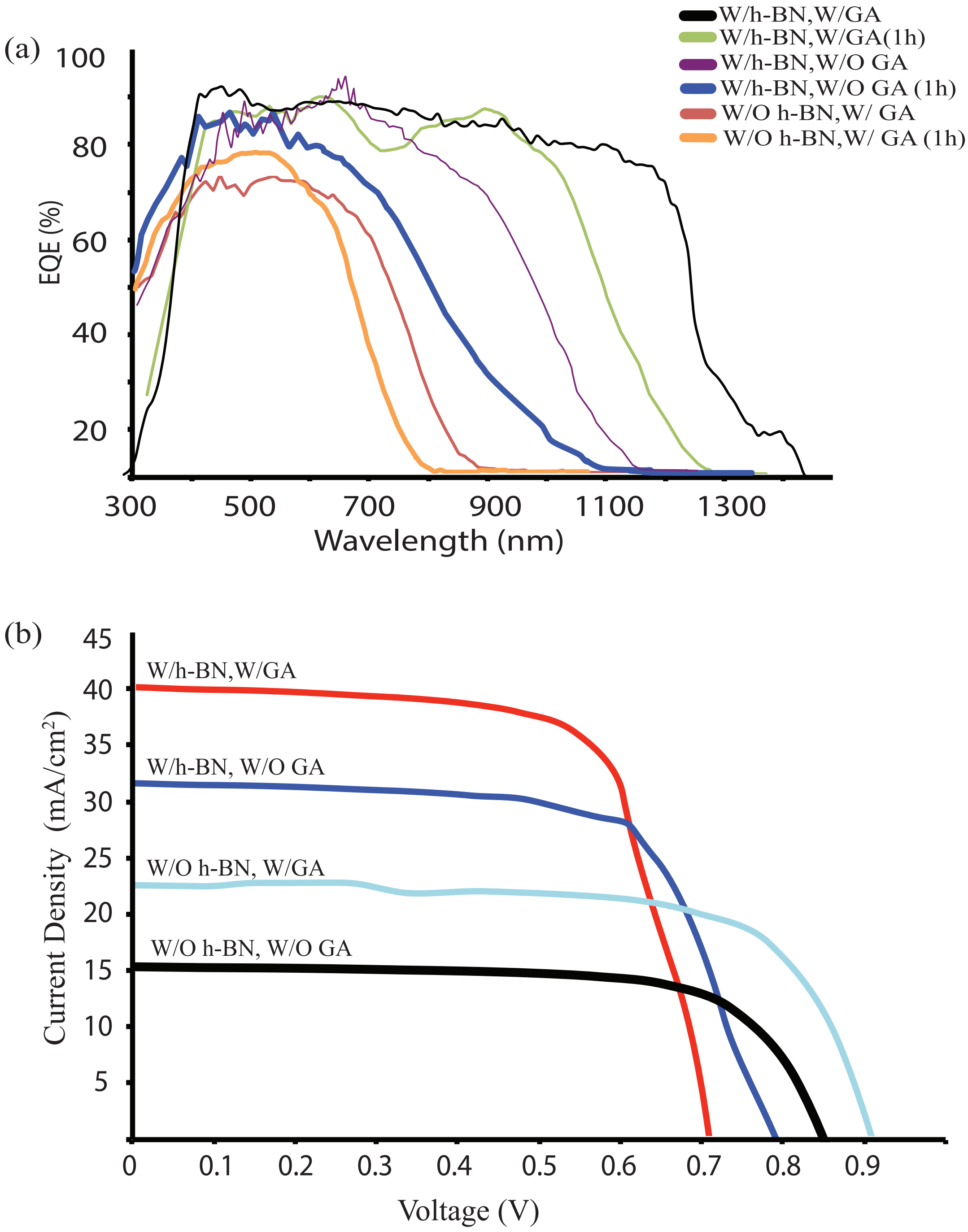

**Figure 3** (a) External quantum efficiency (EQE) spectra for typical graded band gap perovskite cells with and without h-BN and GA components. The cells without GA, display compositional fluctuation due to ionic motion (b) Current density vs voltage characteristics of perovskite solar cells under 1000W/m$^2$ AM 1.5 illumination, showing cells with both h-BN and GA, with h-BN but without GA, without h-BN and with GA, and without both h-BN and GA.

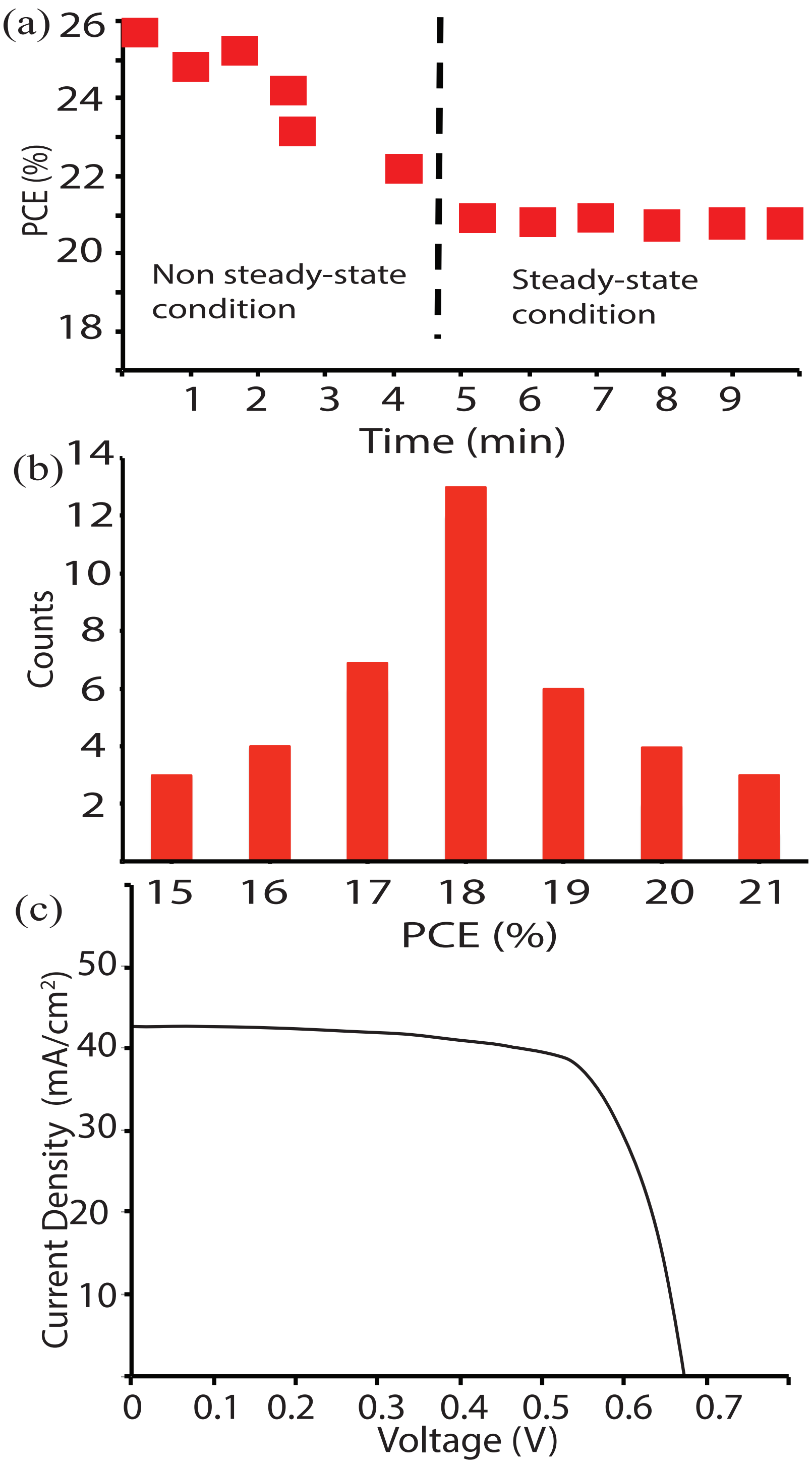

**Figure 4** (a) Time dependence of power conversion efficiency (PCE) for a given graded band gap perovskite cell; the vertical dashed line indicates the onset of steady state behavior. Freshly illuminated cells display PCEs of nearly 26%, for a short period of time. After the cells reach the steady state, they exhibit stable performance at a slightly lower PCE. For this cell the steady state PCE is 20.8%. (b) Histogram of 40 graded band gap solar cells. All PCEs are calculated in steady state. (c) J-V characteristic of a 21.7% PCE cell in the steady state, without antireflective coating.